\documentclass[sigconf]{acmart}
\AtBeginDocument{%
  }

\usepackage{cleveref}
\usepackage{booktabs}
\usepackage{multirow}
\usepackage{enumitem}
\usepackage{tabularx}

\setcopyright{acmlicensed}
\copyrightyear{2026}
\acmYear{2026}
\acmDOI{XXXXXXX.XXXXXXX}
\acmConference[TSMO '26]{Workshop on Two-sided Marketplace Optimization: Search, Discovery, Matching, Pricing \& Growth}{August 10th 2026}{Jeju, Korea}
\acmISBN{978-1-4503-XXXX-X/2018/06}  

\begin{document}

\title{Improving Item Discoverability in e-Commerce Search via Related Intent Generation}

\author{Ji Xin}
\authornote{Equal contribution}
\affiliation{%
  \institution{Instacart}
  \city{Toronto}
  \country{Canada}}
\email{ji.xin@instacart.com}

\author{Xiao Xiao}
\authornotemark[1]
\affiliation{%
  \institution{Instacart}
  \city{San Diego}
  \country{USA}}
\email{xiao.xiao@instacart.com}

\author{Ishan Bhatt}
\affiliation{%
  \institution{Instacart}
  \city{San Francisco}
  \country{USA}}
\email{ishan.bhatt@instacart.com}

\author{Vinesh Gudla}
\affiliation{%
  \city{San Francisco}
  \country{USA}}
\email{vinesh.gudla@gmail.com}

\author{Trace Levinson}
\affiliation{%
  \institution{Instacart}
  \city{Brooklyn}
  \country{USA}}
\email{trace.levinson@instacart.com}

\author{Raochuan Fan}
\affiliation{%
  \institution{Instacart}
  \city{San Francisco}
  \country{USA}}
\email{raochuan.fan@instacart.com}

\author{Shishir Kumar Prasad}
\affiliation{%
  \institution{Instacart}
  \city{San Francisco}
  \country{USA}}
\email{shishir@instacart.com}

\author{Prakash Putta}
\affiliation{%
  \institution{Instacart}
  \city{San Francisco}
  \country{USA}}
\email{prakash.putta@instacart.com}

  \author{Tejaswi Tenneti}
\affiliation{%
  \city{San Francisco}
  \country{USA}}
\email{tentejswi@gmail.com}

\renewcommand{\shortauthors}{Xin et al.}

\begin{abstract}
Traditional search systems are optimized to retrieve items that strictly match a query, often prioritizing precision over recall. In e-commerce marketplaces and particularly grocery, this paradigm is limiting, as user satisfaction and commercial outcomes depend heavily on the discoverability of substitute, complementary, and thematically related items. In this paper, we present a scalable system for \textit{discovery-augmented search} that leverages intent-conditioned recall expansion. Our approach generates implicit user intents to expand candidate recall while maintaining relevance.

The system addresses the cost-quality tradeoff of generative retrieval through a two-stage hybrid architecture. First, we leverage closed-weight large language models (LLMs) to maximize discoverability for head queries. To extend these benefits to tail queries, we then introduce a finetuned small language model (SLM), trained via LoRA adapters and teacher-student distillation. We evaluate the system using a rigorous dual framework: (a) LLM-as-a-judge metrics validated against human preferences for semantic quality, and (b) end-to-end session-level purchase analysis. Results demonstrate that our approach improves both intent generation quality and downstream retrieval effectiveness, extending discovery coverage from approximately 60\% to 80\% of query traffic at roughly 30\% of the teacher model's inference cost, offering a viable path for deployment in large-scale marketplaces.
Beyond relevance gains, discovery-augmented search may serve as a marketplace-balancing mechanism, giving long-tail and emerging supply an opportunity for query-conditioned exposure.

\end{abstract}



\maketitle

\section{Introduction}
Search systems have traditionally been designed around the principle of \textit{ad hoc} retrieval, with the primary goal of returning the most directly matching results for a user’s query. While techniques such as query rewriting and expansion are widely used, they are typically optimized to address vocabulary mismatch and improve matching accuracy for the explicit query, rather than \textit{broadening} the scope of the result set.

In grocery and general e-commerce marketplaces, this strictly relevance-based paradigm is insufficient. Users frequently shop with latent objectives that extend beyond the explicit keyword: seeking reasonable substitutes for out-of-stock goods, gathering ingredients for a specific recipe, or finding complementary items for a use case. In these scenarios, relationships such as \textit{substitution} (e.g., tangerine for clementine), \textit{complementarity} (e.g., pasta and sauce), and \textit{thematic association} (e.g., seafood platter and seasoning) are central to the user's goal but rarely expressed in a single query. Consequently, traditional retrieval under-exposes large portions of the catalog that could satisfy users' broader, implicit intents, thereby limiting revenue potential. In a two-sided marketplace, this under-exposure also disproportionately affects long-tail and emerging supply, framing discovery as a marketplace-balancing concern in addition to a user-satisfaction one.

This paper formalizes a distinct task setup: \textit{discovery-augmented search}. Unlike standard retrieval, the goal here is to retrieve not only items that exactly match the query but also items usefully related through broader semantic associations. We approach this by modeling \textit{implicit intent generation}: we infer users' latent intents using Large Language Models (LLMs) generation and use these generated intents to expand the recall set.

A core challenge in this domain is evaluation where standard relevance metrics (like NDCG on exact matches) often penalize discovery-oriented results. To address this, we conduct analyses using both a novel end-to-end evaluation dataset derived from session co-purchases and an LLM-as-a-judge framework. This dual approach allows us to validate that our generated intents are not only semantically sound but also commercially utility-preserving.

The contributions of this paper are as follows:
\begin{itemize}
    \item \textbf{Task Formulation.} We define \textit{discovery-augmented search} and operationalize it via three latent intent types (substitute, complement, thematic), distinguishing it from query expansion (lexical) and item-item recommendation (query-free).
    \item \textbf{Production-Grade Hybrid Architecture.} A two-tier system that uses cached closed-weight LLM annotations for head queries and a LoRA-finetuned 30B SLM for tail queries, balancing quality and cost at 80+\% query coverage.
    \item \textbf{Dual Evaluation Methodology.} A combined session-derived purchase-prediction benchmark and a human-validated LLM-as-judge framework, providing both extrinsic utility and intrinsic semantic signal in a regime where standard relevance metrics under-credit discovery.
\end{itemize}

\section{Background and Related Work}

\paragraph{Query Understanding and Rewriting}
Query rewriting and expansion techniques~\cite{10.1145/1135777.1135835,10.1145/2071389.2071390,wang-etal-2023-query2doc} have been foundational in web search to improve recall and robustness to lexical mismatch.
In the e-commerce domain, where users frequently reformulate search terms, prior work has leveraged these refinement patterns for improved query understanding~\cite{10.1145/3397271.3401065}.
However, these approaches typically focus on paraphrases, synonyms, and spelling variants, effectively optimizing within the bounds of the user's \textit{explicit} intent. They are generally evaluated on strict relevance metrics (e.g., NDCG, MRR) and do not account for latent intent expansion or the complementarity structures that are critical for discovery for grocery.

\paragraph{Complementarity and Co-Purchase Modeling}
Prior work in recommender systems has extensively explored complementary item prediction using co-occurrence statistics, graph-based methods, and session-based embeddings~\cite{10.1145/2783258.2783381,7738886,10.1145/3219819.3219890,nguyen2025enhancing}.
These methods are effective for item-to-item recommendation surfaces (e.g., "You might also like") but face two limitations in a query-conditioned retrieval setting: they are typically decoupled from the search index and lack query-conditional logic, and they rely on historical co-interaction signals that are sparse for tail queries, emerging supply, and thematic intents (e.g., \textit{smoothie station}) that seldom surface from purchase co-occurrence alone.
Our work is complementary rather than competing: language models infer complementary and thematic relationships directly from the query, extending naturally to long-tail queries where co-purchase graphs are underpopulated, and their outputs can be combined with graph-based signals where dense engagement data is available.

\paragraph{LLMs for Query Understanding and Retrieval}
LLMs have demonstrated efficacy in intent extraction, classification, and attribute extraction~\cite{10.1145/3583780.3615210,electronics14101930,10.1145/3589335.3648298}.
Furthermore, reinforcement learning techniques have been applied to fine-tune LLMs for customized reward signals~\cite{10.1145/3583780.3615474,lin2025recr}.
However, deploying proprietary models across the entire query distribution—particularly the long tail—remains cost-prohibitive.
We address this by transferring reasoning capabilities to cost-efficient models, enabling scalable intent generation without the operational overhead of massive LLMs.

\paragraph{Discovery in Two-Sided Marketplaces}
Search and ranking in two-sided marketplaces must jointly satisfy consumer relevance and supply-side exposure for hosts, sellers, and merchants. Prior production work on Airbnb~\cite{grbovic2018realtime,abdool2020managing,haldar2020improving} explicitly trades off guest preferences against host diversity and listing exposure, and methodological work has formalized seller-side evaluation as a counterfactual problem distinct from consumer-side A/B testing~\cite{hathuc2020counterfactual}. Multi-sided settings such as food delivery further generalize the objective across heterogeneous suppliers~\cite{wang2022recommending}. Discovery-augmented search has analogous implications for both sides of the grocery marketplace: consumers gain access to substitutes and complements they did not know to query for, and retailers and brands---particularly long-tail and emerging ones---have an opportunity to earn query-conditioned exposure beyond exact-match retrieval. Implicit intent generation thus operates as a marketplace-balancing mechanism rather than purely a relevance lever.

\begin{figure}
    \includegraphics[width=\columnwidth]{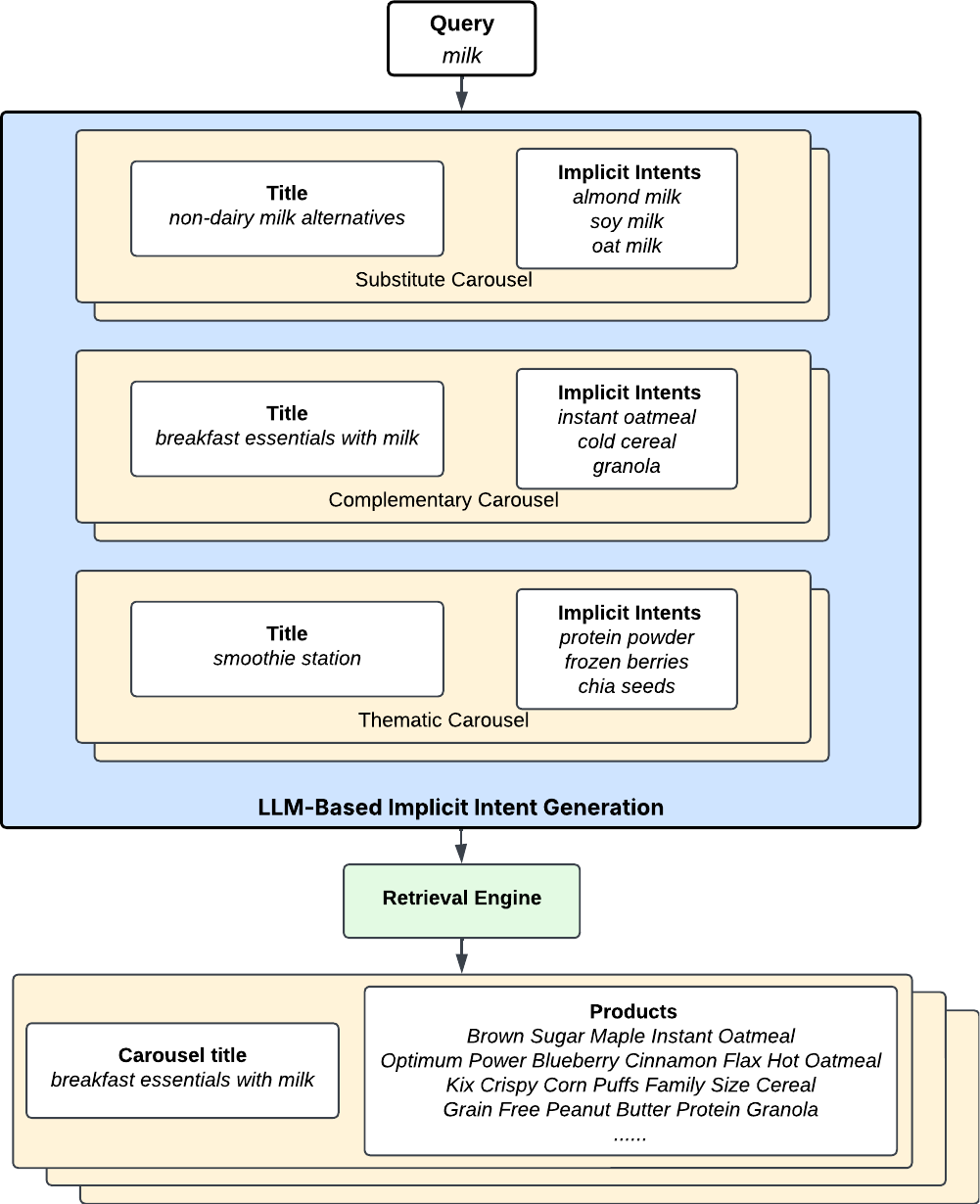}
    \caption{Overview of the implicit intent generation pipeline.
    Example output shown for the query \textit{milk}.}
    \label{fig:overview}
\end{figure}

\section{Implicit Intent Generation for Discovery-Augmented Search}

We introduce the task of discovery-augmented search and discuss our two-stage approach of deploying implicit intent generation to the production system. \Cref{fig:overview} presents an illustration for the end-to-end pipeline. Concretely, the figure walks through the head query \textit{milk}: the LLM emits the carousel title \textit{breakfast essentials with milk} together with intent terms (\textit{instant oatmeal}, \textit{cold cereal}); each term is then dispatched to the existing retrieval engine, whose results hydrate the carousel for display.

\subsection{Problem Formulation}
\label{sec:formulation}

\textbf{Task definition.} Given a user query $q$, the system outputs a set of $C$ carousels $\{(t_i, \mathbf{e}_i)\}_{i=1}^{C}$, where $t_i$ is a natural-language carousel title and $\mathbf{e}_i = (e_{i,1}, \dots, e_{i,K})$ is a length-$K$ list of intent terms. Each $e_{i,j}$ is then issued to the standard retrieval engine to hydrate a product list. The task differs from \textit{query expansion}, which produces a single rewritten query, and from \textit{item-item recommendation}, which is query-free. Concrete values of $C$ and $K$ used in production are $C=9$ and $K=5$.

Given a user query, traditional retrieval systems aim to retrieve products that maximize their relevance.
We instead aim to retrieve an expanded recall set that includes not only highly relevant products but also ones that users implicitly desire. In this paper, we focus on the following latent \textit{intents} of users:
\begin{itemize}
    \item \textbf{Substitute} (e.g., ``non-dairy milk'' for the query ``milk''): a product the user could swap for the queried product.
    \item \textbf{Complementary} (e.g., ``coffee and tea''): a product co-consumed with the queried product.
    \item \textbf{Thematic} (e.g., ``smoothie station''): a product in the same use-case context (recipe, occasion, lifestyle bundle).
\end{itemize}

Our system focuses exclusively on generating these intent terms to populate discovery-oriented \textit{carousels}---interface elements comprising a descriptive title and a curated set of related products---while \textit{exact matches}---products whose tokenized title contains all query tokens after stopword removal and stemming---are handled by the legacy retrieval stack.

\subsection{Closed-Weight LLMs for Head Queries}
\label{sec:head}

The current in-production system uses an offline feature store and covers approximately 10k queries with the highest traffic. These \textit{head} queries cover about 60\% of search traffic while the rest \textit{tail} queries cover 40\%. Annotations for these queries are generated by the closed-weight LLM GPT-3.5 Turbo~\cite{gpt35turbo}.

We utilize a \textit{generate-then-retrieve} paradigm. For each user query, the LLM is instructed to generate a structured output containing two distinct components:
\begin{enumerate}
    \item Carousel Title: A natural language string designed for the user interface (e.g., \textit{breakfast essentials with milk}).
    \item Intent Terms: A list of specific search queries (e.g., \textit{instant oatmeal}, \textit{cold cereal}) that semantically belong to that carousel.
\end{enumerate}
The generated \textit{intent terms} are then executed against the standard search engine to retrieve the final product set, which, along with the \textit{carousel title}, constitutes the carousel for display. This decoupling of generation and retrieval ensures that the LLM focuses on high-level semantic reasoning, while the actual product retrieval remains grounded in the catalog index. We sample with temperature $T=1.0$.

\subsection{Finetuned SLMs for Tail Queries}

To extend discovery capabilities to the long tail (the remaining 40\% of traffic) without incurring prohibitive latency or cost, we introduce a finetuned SLM trained via teacher-student distillation.

We construct a dataset by annotating 20k sampled queries (balanced 50/50 between head and tail) using a stronger teacher model, GPT-5.1~\cite{gpt5}.
To enhance the teacher's reasoning, we enrich the prompts with \textit{contextual metadata} derived from the offline feature store. This includes associated brands, product attributes, conceptual tags, and recent purchase history. We employ few-shot prompting~\cite{NEURIPS2020_1457c0d6} to teach the model to leverage this context when available, or to rely on intrinsic knowledge for cold-start queries.
The resulting dataset is split into training ($70\%$) and validation ($30\%$) sets.
We select Qwen3-30B-Instruct~\cite{qwen3} as our base model. We finetune the model for 1 epoch using LoRA adapters~\cite{lora} with a rank of 8, batch size 65536, and learning rate 1e-4; the standard next-token cross-entropy loss is used over the JSON-formatted teacher outputs.
After finetuning, the SLM is deployed to serve high-frequency tail queries in real time, raising total system coverage from approximately 60\% to 80\% of traffic. The remaining $\sim$20\% of traffic---extreme long-tail queries with low semantic content (e.g., single-character, code-switched, or typo-heavy)---falls through to the legacy retrieval-only stack, without showing any carousels.

\section{Evaluation}

Reflecting the hierarchical nature of our system (Query $\rightarrow$ Title/Intent $\rightarrow$ Product), we adopt a two-tiered evaluation strategy: (1) an end-to-end evaluation of the full pipeline to measure downstream utility, and (2) an intrinsic evaluation of the generation stage to assess semantic quality. Queries selected for evaluation are from the 30\% validation set with no overlap with the training set.

Throughout this section, \textit{Prod.} refers to the currently deployed system: closed-weight GPT-3.5-Turbo annotations cached for the top $\sim$10k head queries (\Cref{sec:head}). By design, Prod. does not cover tail queries; those cells are marked ``—''.

\begin{table}
\caption{Examples from the End-to-End Evaluation Dataset.}
\label{tab:e2e-examples}
\begin{tabular}{ll}
\toprule
Query & Semi-relevant product categories \\
\midrule
chicken broth & Cheese, Fresh Vegetables, Condiments, Pasta \\
tomato paste & Snacks, Cheese, Fresh Fruit, Pasta, Beef \\
tequila & Baking and Cooking, Fruit Juice, Snacks \\
\bottomrule
\end{tabular}
\
\end{table}

\begin{table}[]
\caption{End-to-End purchase prediction results. \textit{Prod.} is the cached GPT-3.5-Turbo head-query system, which does not cover tail queries by design (``—'').}
\label{tab:e2e}
\begin{tabular}{lcccccc}
\toprule
\multirow{2}{*}{Model} & \multicolumn{3}{c}{Head Queries} & \multicolumn{3}{c}{Tail Queries} \\
\cmidrule(l){2-4} \cmidrule(l){5-7}
                       & P & R & F1 & P & R & F1 \\
\midrule
Prod.                  & \textbf{0.130} & 0.260          & 0.173          & --             & --             & --             \\
GPT-5.1                & 0.124          & 0.426          & \textbf{0.192} & 0.108          & 0.379          & 0.168          \\
Qwen3                  & 0.117          & \textbf{0.433} & 0.184          & \textbf{0.117} & \textbf{0.383} & \textbf{0.179} \\
\bottomrule
\end{tabular}
\end{table}

\begin{table}[]
\caption{Alignment between LLM-Judge and Human Experts.}
\label{tab:llm-judge-alignment}
\begin{tabular}{lccc}
\toprule
Metrics   & Title Relevance & Intent Relevance & Intent Novelty \\
\midrule
Precision & 0.84            & 0.87             & 0.95           \\
Recall    & 0.56            & 0.69             & 0.75           \\
F1        & 0.67            & 0.77             & 0.84           \\
\bottomrule
\end{tabular}
\end{table}

\begin{table*}
\caption{LLM Judge scores for all models.}
\begin{tabular}{lllllllll}
\toprule
           & & \multicolumn{4}{c}{Title Metrics} & \multicolumn{3}{c}{Intent Metrics}  \\
           \cmidrule(l){3-6} \cmidrule(l){7-9}
Query & Model & Relevance & Quality & Safety & Coherence & Relevance & Diversity & Novelty \\ \midrule
\multirow{3}{*}{Head}
& Prod.  & 0.941 & 0.983 & 0.995 & 0.964 & 0.911 & 0.917 & 0.853 \\
& GPT-5.1 & 0.967 & 0.995 & 0.996 & 0.977 & 0.966 & 0.968 & 0.898 \\
& Qwen3  & 0.938 & 0.985 & 0.995 & 0.918 & 0.930 & 0.894 & 0.911 \\ \midrule
\multirow{2}{*}{Tail}
& GPT-5.1 & 0.933 & 0.994 & 0.991 & 0.933 & 0.965 & 0.977 & 0.868 \\
& Qwen3  & 0.889 & 0.988 & 0.990 & 0.903 & 0.930 & 0.887 & 0.888 \\
\bottomrule
\end{tabular}
\label{tab:llm-judge}
\end{table*}

\subsection{End-to-End Evaluation}
\label{sec:e2e}

We measure the system's end-to-end performance by constructing a "Discovery Evaluation Dataset" derived from historical search sessions. The construction process involves the following steps:

\begin{itemize}
    \item \textbf{Session Aggregation:} We collect (query, product) pairs where a purchase occurs in the same session as the search. We apply a position-based decay $w(n) = 1 / \log_2(2 + n)$ to the $n$-th purchase in the session (0-indexed, so the first purchase carries weight $1$, the second $\approx 0.63$, and so on), prioritizing earlier purchases as stronger signals of intent.
    \item \textbf{Category Mapping:} Each product is mapped to its high-level product category (e.g., \textit{Mild Gouda Cheese} $\rightarrow$ \textit{cheese} and \textit{Maple Beef Jerky} $\rightarrow$ \textit{snacks}) to improve generalization and reduce sparsity.
    \item \textbf{Signal Extraction:} We aggregate converted products across sessions. To reduce noise as well as to normalize the signals between common vs rare queries, we exclude products that appear outside the top-10 ranking for each query.
    \item \textbf{Exclusion of Exact Matches:} We also exclude products that are exact matches for the query (\Cref{sec:formulation}), as these are handled by the traditional retrieval stack.
\end{itemize}

The resulting dataset represents items that users \textit{implicitly} needed but could not find via exact matching. By removing both exact matches and top-ranked results, the remaining products provide a ground-truth signal for latent needs:\ substitutes, complements, or thematic additions. See \Cref{tab:e2e-examples} for examples. We define a retrieved product as \textit{relevant} for precision and recall iff its high-level category is among the categories purchased in the session, after the position-based weighting above.

We evaluate our models by pairing generated intents with a standard retrieval oracle. \Cref{tab:e2e} compares the performance of our finetuned Qwen3 model against the production baseline (Prod.) and the teacher model (GPT-5.1).
For head queries, both the teacher and student models achieve substantial recall gains at slightly lower precision, translating to a clear F1 improvement over Prod. (0.192 for GPT-5.1 and 0.184 for Qwen3 vs. 0.173). Crucially, on tail queries which the production baseline does not cover, Qwen3 matches or slightly exceeds the much larger GPT-5.1 teacher across all three metrics. This demonstrates that our implicit intent generation approach effectively expands category-level purchase recall for discovery-augmented search while preserving overall retrieval quality.

\subsection{Generation Quality Evaluation}

We evaluate the intrinsic quality of generated carousel titles and intents using an LLM-as-a-judge framework, validated against human preferences. We define metrics at two levels:

\noindent\textbf{Title-level} metrics. For each carousel title $t_i$ produced for a query $q$, the judge returns a binary verdict on:
\begin{itemize}
    \item \textbf{Relevance:} whether $t_i$ is topically aligned with $q$.
    \item \textbf{Quality:} whether $t_i$ is a clear, descriptive natural-language phrase that summarizes the carousel.
    \item \textbf{Safety:} whether $t_i$ avoids sensitive, offensive, or policy-violating content.
    \item \textbf{Coherence:} whether $t_i$ is thematically consistent with the other titles $\{t_j\}_{j \ne i}$ produced for the same query (i.e., the carousel set covers complementary axes rather than duplicating one).
\end{itemize}

\noindent\textbf{Intent-level} metrics. For each intent term $e_{i,j}$ within carousel $i$:
\begin{itemize}
    \item \textbf{Relevance:} whether $e_{i,j}$ semantically belongs under title $t_i$.
    \item \textbf{Diversity:} whether the set $\{e_{i,1}, \dots, e_{i,K}\}$ covers distinct sub-intents rather than near-duplicates.
    \item \textbf{Novelty:} whether $e_{i,j}$ is a non-trivial expansion of $q$ (not a paraphrase or trivial restatement).
\end{itemize}

We employ GPT-5~\cite{gpt5} as the judge model, assigning binary (pass/fail) flags for all metrics. We note a same-provider caveat: the teacher (GPT-5.1) and judge (GPT-5) share a model family, which may bias the judge toward teacher-style outputs; the human-alignment analysis below partially mitigates this concern, and we flag a cross-family judge audit as future work. To validate this automated evaluation, we ask three expert annotators to label a subset of 50 (query, carousel-title, intent-term) tuples on the key metrics: Title Relevance, Intent Relevance, and Intent Novelty.
\Cref{tab:llm-judge-alignment} reports the human--LLM alignment, calculating precision, recall, and F1 between the judge and the majority-vote ground truth across annotators. We observe strong agreement, particularly in precision and F1 scores, confirming that the LLM judge provides reliable signals for model comparison.

It is worth noting that metrics used by the LLM judge are subjective by nature and the LLM judge tends to be stricter than human experts.
For example, for the query \textit{glass bowl}, humans rated the title \textit{meal prep and portion control} as relevant, while the LLM flagged it as irrelevant. Similarly, for \textit{kids lunch}, the intent \textit{grape tomatoes} was marked as novel by humans but not by the LLM.
Consequently, the scores reported in \Cref{tab:llm-judge} likely represent a conservative lower bound on performance.

We show the results by the LLM judge in \Cref{tab:llm-judge}. The teacher model (GPT-5.1) consistently outperforms the production baseline on head queries, particularly in Intent Relevance, Diversity, and Novelty. This indicates that while titles may be of similar quality, the \textit{intents} generated by the LLM retrieve a more diverse and novel set of products.
Our finetuned Qwen3 model successfully retains these benefits, performing comparably to the baseline on most metrics while operating at only $30\%$ of the inference cost.

\begin{table}[]
\small
\caption{Qualitative comparison of carousels generated for the query \textit{cranberry juice}.}
\label{tab:intent-novelty}
\begin{tabularx}{\columnwidth}{@{}p{0.25\columnwidth}@{\hskip 20pt} X @{}}
\toprule
\bf Title                    & \bf Intents                                                                        \\
\midrule
\multicolumn{2}{c}{\bf Prod.}                                                                                 \\
\midrule
Other fruit juices       & orange juice, apple juice, grape juice, pineapple juice, mango juice           \\
\midrule
Healthy alternatives     & pomegranate juice, watermelon juice, acai juice, ginger juice, blueberry juice \\
\midrule
\multicolumn{2}{c}{\bf Qwen3}                                                                                 \\
\midrule
Other cranberry drinks   & cranberry juice cocktail, 100 percent cranberry juice, cranberry apple juice, cranberry grape juice, cranberry flavored sparkling water \\
\midrule
Light and fruity sippers & sparkling water, flavored seltzer, fruit punch, lemonade, sparkling iced tea   \\
\midrule
Office hydration station & mini juice boxes, single serve coffee pods, trail mix, desk water bottle, reusable snack containers             \\
\bottomrule
\end{tabularx}
\end{table}

The most notable improvement seen in \Cref{tab:llm-judge} is \textbf{intent novelty}. \Cref{tab:intent-novelty} illustrates this qualitatively for the query \textit{cranberry juice}. The production baseline restricts itself to strict taxonomic siblings (other juices). In contrast, the finetuned Qwen3 model captures broader thematic intents (e.g., \textit{Office hydration station}), surfacing non-obvious complements like \textit{snack containers} and \textit{coffee pods}. This improved semantic novelty directly correlates with the higher recall observed in the end-to-end evaluation (\Cref{sec:e2e}).

\section{Deployment Lessons and Limitations}
\label{sec:limitations}

Deploying implicit intent generation and evaluating it offline surfaced several failure modes, operational choices, and open caveats that shaped the current design.

\paragraph{Distillation context gap.} The teacher was conditioned on rich offline metadata (brands, attributes, recent purchases). When the student was queried at request time without that metadata payload, quality degraded silently for tail queries. We mitigate by including a controlled fraction of metadata-stripped examples in the distillation set so the student does not rely on the contextual payload.

\paragraph{Coherence sensitivity to decoding.} Carousel \textit{Coherence} (\Cref{tab:llm-judge}) was the metric most sensitive to decoding temperature: high temperatures diversified intents but caused titles to drift off-topic across the carousel set. We selected temperature $T=1.0$ by sweeping on the Coherence metric on a held-out slice.

\paragraph{Long-tail policy.} For the deepest $\sim$20\% of the long tail, neither the cached LLM nor the SLM produced reliably useful intents; we elected to drop carousels for those queries rather than serve degraded results, falling back to retrieval-only.

\paragraph{Maintenance.} Two necessary recurring maintenance tasks are periodically refreshing head- and tail-query annotations as the query distribution and catalog shift, and re-distilling the student model when the teacher is updated or its behavior drifts.

\paragraph{Hallucinated brand names.} Closed-weight LLMs occasionally fabricate plausible-sounding but non-existent brands (e.g., misspellings or invented private labels). Fabricated terms typically return sparse or empty retrieval results downstream, so the affected carousel simply is not visible to the user, but we do not currently apply a pre-retrieval validity check against a curated brand vocabulary.

\paragraph{Counterfactual bias in offline evaluation.} Our end-to-end metric is derived from logged session data, which is itself shaped by the current retrieval and ranking system. Excluding exact matches and top-10 ranked products mitigates this bias by isolating products that users purchased \textit{despite} the system not surfacing them, but does not eliminate it.

\section{Conclusion and Future Work}

We present a scalable framework for improving item discoverability in e-commerce and grocery search through implicit intent generation. By effectively combining the reasoning capabilities of closed-weight LLMs with the cost-efficiency of finetuned SLMs, we demonstrated that it is possible to achieve broad recall expansion across the long tail of the query distribution without prohibitive cost.

Our results suggest that discoverability should be treated as a first-class objective in search systems, particularly in domains characterized by strong complementary structures. Query-conditioned recall expansion provides a flexible mechanism to bridge the gap between precise retrieval and exploratory recommendation, keeping results anchored to the user's immediate context.

Two meta-lessons from deployment may generalize beyond this system. First, decoupling generation from retrieval kept the LLM out of the user-facing hot path and made caching, fallback, and rollout tractable. Second, the dual evaluation framework---session-derived purchase prediction plus a human-validated LLM judge---offers a template for other discovery surfaces where standard relevance metrics under-credit exploration.

This work opens several avenues for future exploration:
\begin{itemize}
    \item \textbf{Non-LLM baselines.} A head-to-head comparison against item-item co-purchase graphs~\cite{10.1145/2783258.2783381,7738886} and classical query-rewriting approaches, both to quantify the marginal value of LLM-based intent generation and to design hybrid pipelines that fall back to graph-based signals where engagement data is dense.
    \item \textbf{Online experimentation.} Large-scale A/B testing to quantify the impact on basket size, session-level revenue, and long-tail supply exposure, providing an online counterpart to our session-derived offline metric.
    \item \textbf{Jointly optimizing} intent generation and product ranking objectives beyond the two-stage pipeline.
    \item Using \textbf{reinforcement learning} to update SLM weights, leveraging online user feedback (clicks, adds-to-cart).
    \item Further \textbf{aligning} offline evaluation datasets and LLM-as-a-judge prompts to correlate with online business metrics.
    \item Generating \textbf{personalized} intents by incorporating user history and session information.
\end{itemize}

\newpage
\bibliographystyle{ACM-Reference-Format}
\bibliography{sample-base}

\end{document}